# Polarimetric Sensitivity of Light-Absorbing Carbonaceous Aerosols Over Ocean: A Theoretical Assessment

Chenchong Zhang[a], William R. Heinson[b], Michael J. Garay[c], Olga Kalashnikova[c], and Rajan K. Chakrabarty[a,*]

[a]*Center for Aerosol Science and Engineering,* Washington University in St. Louis, St. Louis, MO 63130, USA
[b]*Earth System Science Interdisciplinary Center (ESSIC), University of Maryland, College Park, MD 20740 USA, and Climate and Radiation Laboratory, NASA Goddard Space Flight Center, Greenbelt, MD, USA*
[c]*Jet Propulsion Laboratory, California Institute of Technology, Pasadena, CA 91109, USA*

**\*Corresponding author (chakrabarty@wustl.edu)**

**Abstract**

Visible-light-absorbing carbonaceous aerosols within the boundary layer affect the radiance and polarization states of the radiation at the top of the atmosphere. Remote sensing from suborbital and satellite-based platforms utilizes these radiance and polarization signals to retrieve the key properties of these aerosols. Recent retrieval algorithms have shown a progressive trend toward including multi-angular and multi-spectral polarimetric measurements to produce better retrieval accuracy in comparison to those using measurements based on a single viewing angle. Here, we perform a theoretical investigation of the top of atmosphere (TOA) radiance-related reflectance factor (bidirectional reflectance factor (BRF)) and the two types of polarimetry-related factors (polarized bidirectional reflectance factor (pBRF) and the degree of linear polarization (DoLP)) for different types of atmospheric light-absorbing carbonaceous aerosols as a function of particle size distribution. We selected three polarimetric bands corresponding to those utilized by NASA's Airborne Multiangle SpectroPolarimetric Imager (AirMSPI)—near-UV (470 nm), visible (660 nm), and near-infrared (865 nm)—for our simulations which were performed over ocean surface using the successive order of scattering (SOS) algorithm coupled to a Lorenz-Mie aerosol optics model. The analysis of particle phase matrix elements indicates a close relationship between the angular dependencies of DoLP and associated phase matrix components at the shortest polarimetric band (470 nm). Using Jacobian analysis, we find that the radiance- and polarimetry-related reflectance factors of weakly light-absorbing aerosols, such as brown carbon, are more sensitive to changes in particle size and imaginary refractive index in comparison with those of black carbon, which is strongly light-absorbing. Our results suggest that the DoLP data could be used by future retrieval algorithms for reliably estimating microphysical properties of absorbing carbonaceous aerosols with imaginary refractive index less than 0.4.

1. Introduction

Aerosol particles absorb and scatter solar radiation, thus affecting the earth's atmosphere through direct and semi-direct radiative forcing. The scientific community relies on global climate models and remote sensing techniques to determine the radiative effects of various natural and anthropogenic forcing agents. However, the remote sensing responses to aerosol particles, especially for those carbonaceous species that significantly absorb light at visible to near-infrared wavelengths, are among the least understood aspects of the climate system. Previous studies attributed the large uncertainty concerning the remote sensing signals by carbonaceous aerosols to 1) our poor scientific understanding of the spatial and temporal distribution of the aerosol mass, optical properties, etc., and 2) over-simplified parameterizations of aerosols in climate models [1, 2].

Atmospheric carbonaceous aerosols are made up of two major components, each one in both pure and mixed states: organic carbon (OC) and black carbon (BC). The conventional view has been that OC contributes purely to scattering in the visible solar spectrum, and it thereby offsets the warming effects of BC in the atmosphere. This view has been refuted, and observational evidence has shown that OC emissions contain substantial amounts of light-absorbing organic compounds, optically defined as brown carbon (BrC) [3, 4]. In contrast, BC is operationally defined as carbonaceous material with a deep black appearance caused by the significant, wavelength-independent imaginary refractive index [5]. BC is formed by the incomplete combustion of gaseous hydrocarbons in high-temperature combustion systems (e.g., the flaming combustion phase of wildfires) and is a volume absorber (i.e., the absorption cross-section is proportional to the particle volume) [4]. While BC absorbs strongly throughout the solar spectrum, BrC predominantly absorbs long-UV and short-visible (blue and some green) wavelengths. BrC emission has been linked to the low temperature smoldering combustion phase of wildfires and other biomass burning events. Given the significant different optical properties of BC and BrC particles, accurate quantification of their radiative properties is needed to evaluate the role in physical processes that govern remote sensing responses at the top of the atmosphere (TOA).

Remote sensing from airborne and spaceborne platforms provides us useful information for understanding and retrieving the key properties of atmospheric carbonaceous particles. In the context of carbonaceous aerosols, optical parameters of importance to remote sensing algorithms and climate modelers include absorption and scattering cross-sections, single scattering albedo (SSA), and the associated phase matrices. These parameters have a complex dependence on particle size distribution, shape, and composition (hence, their refractive index), and they are also sensitive to observational directions and spectrums. The radiance and polarization acquired by most remote sensors at a specific viewing angle is the integration result of scattered light from the entire column of the atmosphere [6-8]. However, the number of microphysical parameters required for accurate aerosol quantification largely exceeds the number of observables from these sensors. To provide reliable measurements and estimations of aerosols' climate effects, recent remote sensing techniques have progressively included multi-angular polarimeters (MAPs) in different measurement platforms [9]. The inclusion of MAPs has significantly elevated the accuracy of remote sensing techniques to a new level. The instruments endowed with polarimetric capabilities were detailed in several studies [10-12].

MAPs provide us with opportunities for simultaneous retrievals of aerosol characteristics with enhanced information content. A reliable retrieval algorithm depends on the availability of radiative transfer models that are sufficiently accurate to make use of the full information content measured by a MAP. Such radiative transfer models need to address all atmospheric processes and constituents appropriately [13, 14]. Recent polarimetric sensitivity studies have demonstrated that polarization measurements are strongly affected by the aerosols' microphysical properties, optical depths and their vertical layering [15-17]. The question is to find the optimal configuration of spectral wavelengths and viewing angles, and to choose the radiance and polarization quantities to retrieve the microphysical properties of carbonaceous aerosols. In this study, we introduce a formal approach that includes building up the stratified atmosphere, simulating both radiance and polarization signals at the TOA, and quantitatively assessing the sensitivity of remote sensing signals to the microphysical properties of carbonaceous aerosols. We performed our radiative transfer simulation at three wavelengths – near-UV (470 nm), visible (660 nm), and near-infrared (865 nm) – corresponding to the polarimetric bands of NASA's Airborne Multiangle SpectroPolarimetric Imager (AirMSPI). We describe our approach, including the aerosol model selection and the radiative transfer model employed in section 2. The simulation results of the particle phase matrix elements, the reflectance factors, and their sensitivities to microphysical properties are presented in section 3. We provide a summary of our findings and conclude this work in section 4.

2. Method

**(1) Rayleigh optical depths and phase matrix elements of gas molecules**

We calculated the anisotropic scattering cross section ($C_{R,aniso}$) of the dry atmosphere according to Bodhaine et al. [18] and Tomasi et al. [19]. $C_{R,aniso}$, is the product of the isotropic scattering cross section, $C_R$, and the bulk depolarization factor, $F$. The Rayleigh optical depth, $\tau_R(h)$, at any specified altitude $h$ is calculated from

$$\tau_R(h) = C_{R,aniso} \frac{P(h)N_A}{m_a g(h)}, \quad (1)$$

where $P(h)$ is the local atmospheric pressure, $N_A$ is Avogadro's number, $m_a$ is the molecular weight of air, and $g(h)$ is the acceleration of gravity. The phase matrix components, and a set of spherical harmonic expansion coefficients for this Rayleigh scattering phase matrix were also calculated for the Rayleigh atmosphere [20].

**(2) Particle microphysical properties**

Monomodal lognormal size distributions were assumed for all types of carbonaceous particles. The probability density function, $N(r)$, for the number concentration of a group of particles with radii $r$ is written as

$$N(r) = \frac{n_0}{r\sqrt{2\pi}\ln\sigma_g} \exp\left(\frac{-(\ln r - \ln r_g)^2}{2\ln^2 \sigma_g}\right), \quad (2)$$

where $n_0$ is the particle number concentration (in #/m$^3$), and $r_g$ and $\sigma_g$ are the geometric mean radius and the standard deviation, respectively. We assumed the $\sigma_g$ for all types of carbonaceous aerosols was constant at 1.6. This is a mean estimation of $\sigma_g$ following Wang et al. [21] and Drury et al. [22]. We adjusted $r_g$ to acquire different particle size distributions in this study. The effective diameter, $d_{eff}$, is related to $r_g$ by

$$d_{eff} = 2r_g \exp\left(\frac{5}{2}\ln^2 \sigma_g\right). \quad (3)$$

For this study, we selected seven values of $d_{eff}$. Their associated $r_g$ values are shown in Table 1. The selected $d_{eff}$ values are the lower limits of the first six particle size bins (size 1 to 6) and the upper limits of the sixth particle size bin (size 7) used by the Model for Simulating Aerosol Interactions with Chemistry (MOSAIC), which is widely-used to study the emission properties and impacts of biomass burning [23]. Previous studies assumed the dry carbonaceous particles from primary anthropogenic emissions fall mainly within the size range bounded by these six size bins [24, 25].

Table 2 lists the refractive indices of the different carbonaceous aerosols. For BC particles, we chose a relatively higher absorbing scenario, $BC_{high}$, adapted from Bond et al. [5], and a lower absorbing counterpart, $BC_{low}$, adapted from Hess et al. [26]. Both refractive indices were observed to be consistent with different independent measurements. OC, representing organic carbon with extremely weak light absorption in the atmospheric environment, had the smallest refractive index values for both its real ($RI_r$) and imaginary parts ($RI_i$). We further emphasize that BC and OC's refractive indices were spectrally independent in this study.

In contrast, the imaginary refractive index of BrC follows a Kramers-Kronig dispersion relation for a damped harmonic oscillator at wavelengths shorter than 572 nm, which is given by

$$n_i = a \frac{\gamma v}{(v_0^2 - v^2)^2 + (\gamma v)^2}, \quad (5)$$

where the constants $a$ and $\gamma$ are $10^{29}$ $s^{-2}$ and $2 \times 10^{13}$ $s^{-1}$, respectively [27, 28]. The frequency term $v$ is defined as $c/\lambda$, where $c$ is the speed of light and $\lambda$ is the incident wavelength. The principal frequency, $v_0$, is calculated at a wavelength of $\lambda_0 = 300$ nm. The imaginary part of the refractive index for BrC diminishes to 0.001 at wavelengths longer than 572 nm. For all types of aerosols, we used Mie theory to calculate the absorption cross sections, $C_{A,abs}$, scattering cross sections, $C_{A,sca}$, and scattering phase matrix components, $F_a(\Theta)$, where $\Theta$ represents the scattering angle [29].

**(3) Radiative transfer modeling**

We employed a 1-dimensional vector radiative transfer (VRT) code based on plane-parallel successive-order-of-scattering (SOS) principles to calculate the angular distributions of the polarized radiances scattered by the simulated atmosphere-ocean system [30, 31]. In an SOS model, the directional space is represented discretely as a Fourier series of the azimuthal angle and an angular quadrature for the zenith angle, as illustrated in Fig. S1 [32]. The solar zenith angle, $\theta_0$, and the viewing zenith angle, $\theta_v$, were measured from the vertical z-axis. The solar azimuth angle, $\phi_0$, and the viewing azimuth angle, $\phi_v$, were measured clockwise from the y-axis, which points in the 0° direction. In practice, the relative azimuth angle, $\phi$, which is defined as the angle measured clockwise from $\phi_0$ to $\phi_v$, is the direct input parameter of the SOS code, instead of $\phi_0$ and $\phi_v$.

The atmospheric profile was represented as a sequence of finitely thick and optically uniform layers assembled between the TOA and the oceanic surface. For each stratified layer, the input files to the SOS code contain the optical depths, single scattering albedo SSA, and the scattering phase matrix components. The radiance and polarization acquired by most remote sensing sensors at a specific viewing angle is the integration result of scattered light from the entire column of the atmosphere. Due to the limitation of resolving vertical inhomogeneity of atmospheric columns, the simulated atmosphere is normally prescribed as a single aerosol-mixed layer in the retrieval algorithms [33-35]. In this study, aerosol particles were confined within one optically homogeneous layer extending to 2 km above the oceanic surface following

Kalashnikova et al. [36]. The wind flows over the oceanic surface with a constant speed of 5 m/s. The wave slope distributions were applied under this wind speed assumption [37]. Within the aerosol-mixed layer, the light scattering properties come from a mixture of aerosols and gas molecules. The total optical depth, $\tau_L$, of the mixed atmospheric layer can be calculated by adding the optical depths of the aerosols ($\tau_A$) and the pure Rayleigh atmosphere ($\tau_R$):

$$\tau_L = \tau_A + \tau_R. \quad (6)$$

The optical depth of the aerosols can be obtained from the scattering cross sections and absorption cross sections by

$$\tau_A = N_0 (C_{A,sca} + C_{A,abs}). \quad (7)$$

In this study, the optical depths contributed by different particle models were firstly fixed as 1.0 at 550 nm. As Eqn. 6 shows, the subscripts "$L$", "$A$", and "$R$" denote the parameters for the atmospheric layers, aerosols, and gas molecules, respectively. The SSA of the mixed layer, $\omega_L$, was calculated by

$$\omega_L \cdot \tau_L = \omega_A \cdot \tau_A + \omega_R \cdot \tau_R. \quad (8)$$

Note here SSA of Rayleigh atmosphere ($\omega_R$) equals to unity for all studied wavelengths. The phase matrix components are calculated by

$$F(\theta)_L^j = \frac{\omega_A \tau_A F(\theta)_A^j + \tau_R F(\theta)_R^j}{\omega_A \cdot \tau_A + \tau_R}, \quad (9)$$

where $F(\theta)$ represents $4 \times 4$ matrices of expansion coefficients. The superscript $j$ stands for the $j$-th component in the matrix ($j = 1, 2, \ldots, 16$). To describe the effects of polarization on light scattering at the TOA, we use the traditional Stokes vector and phase matrix formalism:

$$\begin{pmatrix} I_s \\ Q_s \\ U_s \\ V_s \end{pmatrix} = \begin{pmatrix} F_{11}(\Theta) & F_{12}(\Theta) & 0 & 0 \\ F_{12}(\Theta) & F_{22}(\Theta) & 0 & 0 \\ 0 & 0 & F_{33}(\Theta) & F_{34}(\Theta) \\ 0 & 0 & -F_{34}(\Theta) & F_{44}(\Theta) \end{pmatrix} \begin{pmatrix} I_i \\ Q_i \\ U_i \\ V_i \end{pmatrix}. \quad (10)$$

Here the phase matrix transforms the incident light's Stokes vector $[I_i, Q_i, U_i, V_i]^T$ to the scattered Stokes vector $[I_s, Q_s, U_s, V_s]^T$, where $I$ represents the total intensity, $Q$ represents the difference between the horizontally and vertically polarized intensities, $U$ represents the difference between the $+45°$ and $-45°$ polarized intensities, and $V$ represents the difference between the right-handed and left-handed polarized intensities. Incident solar radiation is unpolarized, with a Stokes vector of $(I_i, 0, 0, 0)^T$. The optical properties of the medium transmitting the light are contained in the $4 \times 4$ matrix in Eqn. (10). When the particles are randomly oriented, they have a plane of symmetry, and thus the eight off-axis elements are zero. Furthermore, if particles are spherical, we have two additional coefficient relationships, $F_{11}(\Theta) = F_{22}(\Theta)$ and $F_{33}(\Theta) = F_{44}(\Theta)$. The values of $F_{33}(\Theta)$ and $F_{34}(\Theta)$ indicate the proportion of particles that are nonspherical or aggregated in the particle groups. These two matrix components have negligible influences among different particle models in this study.

The interactions between solar radiation and atmospheric species change the total intensity, $I$, and the polarized intensities, $Q$ and $U$. We note here that the circular polarization, $V$, will always be negligible in atmospheric light scattering. As a result, the total polarized intensity is calculated by

$$I_{pol} = \sqrt{Q^2 + U^2}. \quad (11)$$

In practice, the Stokes parameters measured by airborne and spaceborne sensors are scaled by the incident solar irradiance [38]. These dimensionless reflectance factors include the radiance-related factor, bidirectional reflectance factor (BRF), and two polarization-related factors, the polarized bidirectional reflectance factor (pBRF) and the degree of linear polarization (DoLP). The BRF is a measure of the total reflected radiance $I$, and is given by the expression

$$\text{BRF} = \frac{\pi I}{\mu_0 F_0}, \qquad (12)$$

where $\mu_0$ is the cosine value of the solar zenith angle and $F_0$ is the extraterrestrial solar irradiance. pBRF is defined similarly to BRF but only replacing the total intensity $I$ with the polarized radiance, $I_{pol}$, as

$$\text{pBRF} = \frac{\pi I_{pol}}{\mu_0 F_0}. \qquad (13)$$

The DoLP is the ratio of pBRF to BRF, or $I_{pol}/I$, and is defined as

$$\text{DoLP} = \sqrt{(Q/I)^2 + (U/I)^2}. \qquad (14)$$

The values of these three types of dimensionless reflectance factors are determined by the integrated contribution of the optical depths, SSA, and the phase functions for the mixture of molecules and aerosols within the atmosphere-ocean system. The magnitudes of BRF, pBRF and DoLP at specified scattering angle $\Theta$ are proportional to the $F_{11}(\Theta)$, $F_{12}(\Theta)$, and $-F_{12}(\Theta)/F_{11}(\Theta)$ in Eqn. (10), respectively if the optical properties of the coupled atmosphere-ocean system are homogeneous and the optical depths and SSA are similar for independent forward simulations. But the simulated atmosphere-ocean system in this study contains multiple atmospheric layers with distinct optical properties as well as a rough oceanic surface. The Mueller matrix components of the aerosol models or the aerosol-mixed layer cannot quantitatively explain the values of reflectance factors detected at the TOA. However, the microphysical properties of aerosol models are the only variated input parameters in multiple parallel simulations, the differences of those matrix components will reflect the variations in the reflectance factors. Our sensitivity analysis in Section 3 will discuss how aerosols' microphysical properties affect these reflectance factors.

3. Results

Plots of the independent elements of the phase matrix of aerosols with different refractive indices are shown in Fig. 1 for incident wavelengths $\lambda$ of 470 nm, 660 nm, and 865 nm. The three selected wavelengths are the polarimetric bands of the Airborne Multiangle SpectroPolarimetric Imager (AirMSPI), which is an eight-band pushbroom camera mounted on a gimbal [10]. Phase matrix elements of BC are shown as red lines (solid line for $BC_{high}$ and dashed line for $BC_{low}$), and two types of organic aerosols are shown as blue lines (solid line for BrC and dashed line for OC). The size distributions of four different types of aerosols are identical ($d_{eff}$ = 90 nm, $\sigma_g$ = 1.6), and differences in angular dependencies of the phase matrix elements are attributable to the different refractive indices of the particles. As a reference to particle models, molecular scattering patterns of a Rayleigh atmosphere are also shown in the black dash-dotted lines in all subplots of Fig. 1. The elements of the phase matrix for molecular scattering are spectrally independent at the three studied wavelengths.

Fig. 1 (a)-(c) show that the scattering phase functions, $F_{11}$, of four different types of aerosols are all strongly forward peaked. In contrast, the phase function of a Rayleigh atmosphere is symmetric in the forward and backward directions, with a local minimum at 90 degrees. These three subplots also show that the phase functions of strongly light-absorbing BC decrease monotonically as the scattering angle increases. For weakly light-absorbing BrC and OC, the monotonically decreasing trends continue until the scattering angle reaches 130 degrees. As the scattering angle continuously increases toward the backscattering directions, BrC's and OC's phase functions slightly increase again. These different phase function patterns among strongly and weakly light-absorbing aerosols are more obvious at the shortest wavelengths in this study (470 nm). These results imply that as the mean size parameter, defined as $\pi d_{\text{eff}}/\lambda$, continuously decreases, the phase matrix elements of different types of aerosols tend to converge to similar values.

Fig. 1 (d) to (f) show the polarized phase function $F_{12}$ as a function of the scattering angles. Just like the $F_{11}$ plots, the differences between BC and weakly light-absorbing organic aerosols are larger at 470 nm than those longer wavelengths. Fig. 1 (d) shows that two types of BC aerosols have negative peaks at 30 degrees, while BrC and OC show low positive peaks at a scattering angle of 150 degrees. The positive peaks of BC particles appear at the similar angles and the peak magnitudes hold through all three wavelengths. In contrast, the lower positive peaks at 470 nm for BrC and OC quickly disappear as the incident wavelength increases, and the similar negative peaks of $F_{12}$ appear at the near-forward direction. It turns out that the difference in the $F_{12}$ patterns between strongly and weakly light-absorbing aerosols becomes negligible among all scattering angles at longer wavelengths.

The subplots in the bottom row show the angular dependencies of the degree of linear polarization, defined as the $F_{11}$ normalized ratio, $-F_{12}/F_{11}$. This ratio ranges from

-1 to 1 for all kinds of particles and molecules. The scattered light of a pure Rayleigh atmosphere is 100% polarized (-$F_{12}/F_{11}$ equals 1) at a scattering angle of 90 degrees. BC aerosols ($BC_{high}$ and $BC_{low}$) show maximum DoLP values of 0.33 at 470 nm, 0.54 at 660 nm, and 0.68 at 865 nm near a scattering angle of 90 degrees. For BrC and OC aerosols, the positive maximums of this ratio are smaller. Also, these two types of weakly light-absorbing aerosols both show extra negative DoLP peaks at a scattering angle around 150 degrees for the two shorter incident wavelengths. The negative DoLP peaks, which are the signature of weakly light-absorbing aerosols, disappear at an incident wavelength of 865 nm. Besides, the magnitudes of the positive DoLP peaks for BrC and OC are also much closer to the BC's peak at this longer wavelength as the angular dependency of $F_{12}$.

The phase matrix elements of the aerosol-mixed layer are shown in Fig. S2. The curves indicate the angular dependencies of phase matrix elements introduced above are conserved after mixing with a Rayleigh atmosphere. Fig. 1 and Fig. S2 demonstrate that the most significant differences in phase matrix elements among different types of aerosols appear at the shortest wavelength ($\lambda$=470 nm). In particular, the different patterns of $F_{11}$ and -$F_{12}/F_{11}$ appear when the scattering angle exceeds 90 degrees, which the viewing geometries of the real-world sensors can cover. The calculated phase matrix components are then applied into the forward radiative transfer model. The viewing geometries are defined as: the solar zenith angle, $\theta_0$, is fixed at 30 degrees as a representation of mid-latitude illumination geometry. The chosen viewing zenith angles, $\theta_v$, have a range of $\pm 68°$, which are consistent with the AirMSPI sensors [39, 40]. The relative azimuth angle, $\phi$, is sampled between $0°$ and $180°$. We calculated the scattering angle $\Theta$ through

$$\cos(\Theta) = -\cos(\theta_0) \cdot \cos(\theta_v) + \sin(\theta_0) \cdot \sin(\theta_v) \cdot \cos(\phi). \quad (15)$$

We selected the different viewing geometries to acquire a series of increasing scattering angles. Table 3 lists the detailed observational geometries for further forward radiative simulations in this study. Given the angle ranges introduced above, the resulting scattering angles are between 81.6 degrees and 177.9 degrees.

Fig. 2 shows the results of the reflectance factors at the TOA for a Rayleigh scattering atmosphere containing an aerosol-mixed layer from the oceanic surface to 2 km. The AODs of different types of aerosols are fixed at 1.0 at λ equals to 550 nm. Two types of BC aerosols have similar SSA values and phase matrix elements for three studied wavelengths. These similarities are also observed in BrC and OC models. As a result, the reflectance factors of $BC_{high}$ (red squares) and BrC (green circles) approximately overlap those of $BC_{low}$ (black solid lines) and OC (blue dashed lines), respectively. As shown in the first and second rows of subplots in Fig. 2, both BRF and pBRF for BrC and OC aerosols are larger than those for BC aerosols because the weakly light-absorbing aerosols normally have higher SSA values. The radiation energy dissipated in the aerosol-mixed layer is always smaller for those aerosols with lower light absorption. SSA values partly affect the reflectance factors at different wavelengths. As the wavelength increases from 470 nm to 865 nm, SSA values for the same type of BC aerosol decrease. This decrease directly leads to their monotonically descending BRF values at the same scattering angles in Fig. 2 (a), (b) and (c). Besides the SSA effects, we also observed that weakly light-absorbing BrC and OC aerosols show a minimum BRF at a scattering angle around 150 degrees in Fig. 2 (a), which is consistent with the minimum of $F_{11}$ at the same angle in Fig. 1 (a). It implies the particle phase matrix elements have the dominant influence on the angular dependency of BRF for all types of particles.

Fig. 2 (d), (e), and (f) demonstrate the angular and spectral dependencies of pBRF. All types of aerosols show a significant decreasing trend of pBRF as the scattering angle increases. Fig. 1 (d), (e), and (f) have shown the magnitudes of $F_{12}$ for four kinds of aerosols are all close to zero at a scattering angle larger than 80 degrees. Thus the angular dependency shown in the subplots in the second row of Fig. 2 is mainly determined by the monotonically decreasing magnitude, $|F_{12}|$, of Rayleigh scattering by gaseous molecules. As the incident wavelength shifts from 470 nm (Fig. 2 (d)) to 865 nm (Fig. 2 (f)), the magnitudes of pBRF for BC particle models show a weak decreasing pBRF trend. We note here that the scattered light detected at the TOA is always partially polarized, and pBRF is a measure of the relative intensity of that polarized portion. The spectral dependencies of pBRF between BC and BrC, as well as BC and OC aerosols, implies the SSA values have different importance to the magnitudes of polarized radiance among variated types of aerosols. Fig. S3 shows the spectral dependency of the values of SSA for four types of aerosols at all eight wavelength bands of AirMSPI. SSA values of two types of BC aerosols show a decreasing trend as the incident wavelength increases. In contrast, SSA of BrC and OC remain constant as the wavelength exceeds 470 nm. This result is consistent with the spectral dependency of pBRF we introduced above.

Subplots in the bottom row of Fig. 2 show the DoLP values at the three studied wavelengths. DoLP is the ratio of pBRF to BRF. Unlike the two reflectance factors introduced above, strongly light-absorbing aerosols show higher DoLP at the three studied wavelengths with less significant spectral dependencies. It implies that the scattered light from BC particles contains a larger portion of polarized radiation in the total radiation. Also, the spectral dependencies appeared in BRF and pBRF offset each other for BC particles. In contrast, BrC and OC conserve the increasing DoLP magnitudes as the incident wavelength shifts to the near-infrared regime (865 nm). Like the relationship between the BRF and $F_{11}$ plots, we find the normalized phase matrix elements $-F_{12}/F_{11}$ determine the angular DoLP patterns. We have observed similar peaks at a scattering angle of 90 to 100 degrees as shown in Fig. 1 (g), (h), and (i).

Fig. 2 and Fig. S3 show that phase matrix elements and SSA affect the spectral and angular dependency of reflectance factors. Fig. 2 (d) to (i) demonstrate that the reflectance factors are sensitive to aerosol type, specifically the imaginary refractive indices, at a scattering angle between 80 and 140 degrees. The decreasing sensitivity of reflectance factors to the aerosol type at larger scattering angles is because of the low magnitudes of the polarized radiation for all types of aerosols. The aerosol microphysical properties in future work should be retrieved in the angle range smaller than 140 degrees for the higher sensitivity and signal-to-noise ratio purposes.

We further extended our simulations to different AOD values in the boundary layer to study the sensitivities of associated remote sensing signals to the aerosol mass concentration. Given the significant similarities between the two BC species, and the BrC/OC species, we selected $BC_{high}$ and BrC as the representatives of strongly and weakly light-absorbing particles, respectively. Fig. 3 (a) to (f) show that the reflectance factors of both BC and BrC particles conserve the angular dependencies shown in the unit AOD circumstance. The comparisons between Fig. 3 (a) and (d), as well as (c) and (f) illustrate that the magnitudes of BrC aerosol BRF and DoLP are more sensitive to the AOD levels than those for BC particles. It is because the necessary change of the BC mass concentration to achieve a unit variation of layer AOD is less than that for BrC particles. It directly leads to a less significant modulation to the atmospheric radiometric and polarimetric signals.

Besides the refractive index, the optical properties of atmospheric particles are also affected by their size. In the following sections, we will incorporate variated size distributions into the sensitivity analysis of particle properties. Fig. 4 shows the influences of $d_{eff}$ on the phase matrix elements at an incident wavelength of 470 nm. Detailed information about $d_{eff}$ for sizes 1 to 7 is listed in Table 1. Previous analysis shows BrC and OC at 470 nm are optically similar. We just show the matrix elements of BrC in Fig. 4. Like Fig. 1, we also plot the matrix elements of the reference Rayleigh atmosphere (black dashed-dotted lines) in each subplot. Fig. 4 (a), (b), and (c) show that with increasing particle size, the plot for $F_{11}$ evolves from forward and backward symmetry to a strongly forward-peaked shape. The peaks in the forward directions reach a maximum for aerosols with the largest $d_{eff}$. The phase functions $F_{11}$ of BC (Fig. 4 (a)) near the backscattering angles show a monotonically decreasing trend as the particle size increases for all studied sizes. In contrast, when $d_{\text{eff}}$ exceeds 39.1 nm, $F_{11}$ of BrC (Fig. 4 (c)) distinctly increase in the near-backscattering directions. As a result, a minimum $F_{11}$ value is observed at about 140 degrees.

Fig. 4 (g), (h), and (i) show the effects of particle size on the ratio $-F_{12}/F_{11}$. As $d_{\text{eff}}$ increases from 39.1 nm to 156.0 nm, the $-F_{12}/F_{11}$ peaks of the three types of particles decrease. These peaks all appear at a scattering angle slightly larger than 90 degrees. For two types of BC particles, as the particle sizes continue increasing, the magnitudes of the DoLP peaks increase again to a maximum of over 0.9. Furthermore, the peak of BC particles shifts to the forward scattering directions. In contrast, the positive DoLP peaks of BrC completely disappear as the particle geometric mean radii exceed 156.0 nm. Instead, a negative DoLP peak appears at a scattering angle of 150 degrees. For those particles with $d_{\text{eff}}$ larger than 313 nm, the $-F_{12}/F_{11}$ ratios are less angularly dependent, until the scattering angle approaches the backscattering directions. We highlight the angular dependencies of the related phase matrix components for a scattering angle range between 80 degrees and 180 degrees in Fig. S4. The phase component values in this angle range are closely related to the further reflectance factor computations.

The solid lines within the shaded area in Fig. 5 show the mean values of the simulated reflectance factors. The angular dependencies of the mean reflectance factors are consistent with the corresponding factors shown in Fig. 4 (a), (d), and (g). The width of the shaded area represents the value ranges of the reflectance factors as the particle geometric mean radius increases from 39.1 nm to 2500.0 nm. The different area widths of BrC and BC aerosols indicate the particle size more significantly affects the reflectance factors of weakly light-absorbing particles than those for BC particles. In Fig. 5 (c), we have observed the similar decreasing sensitivities of DoLP to the effective diameters at near-backscattering directions for all kinds of aerosols. It is caused by the extremely low magnitudes of the polarized radiation for all types of aerosols (as shown in Fig. 5 (b)) within this angle range. The extra comparisons of the reflectance factors between OC and two types of BC aerosols are shown in Fig. S5. The effects of size distributions on associated reflectance factors of OC are similar to those for BrC.

Fig. 1 to Fig. 5 show that both refractive index and effective diameter affect the phase matrix components and resulting reflectance factors at the TOA. This effect is prominently significant at shorter wavelengths and at scattering angles between 80 and 140 degrees. We also notice that the differences in the imaginary refractive index values of $BC_{high}$ and $BC_{low}$, as well as $BC_{low}$ and BrC, are over 0.3. However, the optical dissimilarities between $BC_{high}$ and $BC_{low}$ are not so obvious as those between $BC_{low}$ and BrC even if we consider the effects of particle size. As a result, we expanded the levels of the imaginary refractive indices, $RI_i$, to an entire range between 0 and 1 with a step size of 0.1 to quantify the sensitivity of the studied reflectance factors to particle absorbing capacities and effective diameter simultaneously. The real parts of the refractive indices were fixed at 1.6. We also applied a cubic spline interpolation method to draw the 2-dimensional contour plots. Fig. 6 illustrate the magnitudes of BRF (Fig. 6 (a)) and pBRF (Fig. 6 (b)) at a scattering angle of 81.6 degrees and an incident wavelength of 470 nm. As shown in Fig. 6 (a), BRF for any fixed effective diameter monotonically decreases as the imaginary refractive index increases. This pattern holds for all the studied effective diameters when imaginary refractive index was less than 0.1. According to Table 2, BrC and OC aerosols both fall within this $RI_i$ range. Conversely, the sensitivity of BRF to $RI_i$ dramatically decreases when $RI_i$ continuously increases.

Unlike BRF, pBRF in Fig. 6 (b) shows much lower sensitivities to $RI_i$ and effective diameters. The polarized phase function ($F_{12}$) plots in Fig. 1 and Fig. 4 indicate that the polarized portion of the total reflected radiance is mainly contributed by the scattering of the Rayleigh atmosphere. It results in less varied pBRF magnitudes at the TOA. Combining two contour plots in Fig. 6, we also found that the pBRF almost reaches the magnitudes of total reflected radiance for those strongly light-absorbing particles as BRF continues to decrease. The increase in polarization can also be directly observed in the DoLP contour plot (Fig. 7 (a)). The high magnitudes of DoLP generally appear in the high $RI_i$ region. Those large DoLP values result from the collective effects on polarization between a Rayleigh atmosphere and strongly light-absorbing aerosols. DoLP of these aerosols initially decreases and then increases with an increase in $d_{eff}$. It is consistent with the changing patterns of -$F_{12}/F_{11}$ as shown in Fig. 4 (g) and (h). For weakly light-absorbing aerosols, the decreasing trend of DoLP with the increase of $d_{eff}$ holds for small particles. As the $d_{eff}$ continuously increase, DoLP is nearly constant for a wide size range. This result is consistent with the -$F_{12}/F_{11}$ ratios we illustrated in Fig. 4 (i) at a scattering angle of 81.6 degrees. The magnitudes of -$F_{12}/F_{11}$ for large BrC aerosols (shown as the orange and yellow lines in Fig. 4 and Fig. S4) are comparable to each other, and significantly lower than those of smaller BrC particles (shown as the dark red lines). We have also shown the contour plots of DoLP magnitudes at 99.81 and 144.71 degrees in Fig. S6 (a) and (b), respectively. Those two figures indicate the sensitivities of DoLP to particle size and refractive index gradually diminish as the scattering angle increases. The further retrieval algorithms built on the similar predefined atmospheric conditions should utilize the polarized reflectance factors at a scattering angle close to 90 degrees.

Sensitivity of DoLP to imaginary refractive index $RI_i$ is quantified through the finite-difference form of Jacobians, $|\delta DoLP/\delta RI_i|$, in Fig. 7 (b). For all studied particle size distributions, Jacobians of the DoLP show prominently high values for particles with extremely weak light absorption ($RI_i$ less than 0.1). The inset of Fig. 7 (b) shows the Jacobians of DoLP for two size distributions with the smallest and largest effective diameters as $RI_i$ exceeds 0.1. It proves the DoLP sensitivities to $RI_i$ are also affected by particle sizes. Small particles have larger sensitivity for the relative change of $RI_i$ values when $RI_i$ is less than 0.4. The different sensitivities indicate that, in the retrieval algorithm, the measurements of polarization contain more size information for $RI_i$ values less than 0.4.

DoLP is the parameter used to assess the required accuracy of a MAP since the selection of the polarimetric reference plane has little influence on this metric in practice [41]. The Decadal Survey ACE requirement for polarimetric uncertainty is 0.5% (0.005 alternatively). For most scene reflectance, the uncertainties are below 0.01 [42]. The spectral bandwidth of the AirMSPI instrument centered at 470 nm is 45 nm [43].

A quantitative analysis of the polarimetric reflectance factors within this bandwidth are important to evaluate the instrument accuracy and retrieval quality. Fig. 8 shows the relative deviation of DoLP of $BC_{high}$ and BrC particles at a scattering angle of 81.6°. We define the relative deviation of DoLP between certain wavelength $\lambda$ and the centered wavelength (470 nm) as $|DoLP_{\lambda,d_{eff}} - DoLP_{470nm,d_{eff}}|/DoLP_{470nm,d_{eff}}$. The width of the yellow area in Fig. 8 represents the wavelength range which satisfies the measurement uncertainty requirements. We call it "acceptable range" in the following discussion. $BC_{high}$ particles with a mean effective diameter larger than 313 nm show a wider acceptable range than the small-sized counterparts as the wavelength changes from the center to the edges of the bandwidth. In contrast, the acceptable range of BrC particles monotonically decreases as the size increases. It implies that the DoLP measurement provides more reliable results for those large-sized BC particles and small-sized BrC particles at the 470 nm band when we take the finite spectral bandwidth into consideration.

## 4. Summary and discussions

Light-absorbing aerosols in the boundary layer sensitively affect the radiative transfer phenomena in the atmosphere. The scattering and absorbing effects of aerosols simultaneously modify the intensities and polarization states of the radiation. Photo-polarimetric technique offers an effective solution to retrieve the complex microphysical properties of aerosols. The first step of the retrieval method is to generate the dataset of reflectance factors with the input of a wide-range of aerosol microphysical properties. In this study, we integrated the Mie code with the SOS algorithm to simulate the reflectance factors at the TOA of an atmosphere-ocean system. The computation through Mie code shows weakly light-absorbing BrC and OC have distinct phase function ($F_{11}$) and degrees of linear polarization ($-F_{12}/F_{11}$) patterns compared to those for strongly light-absorbing BC aerosols. The difference is more significant at an incident wavelength of 470 nm than two longer counterparts. $F_{11}$ of BC aerosols monotonically decreases with increasing scattering angles within an angle range between 90 and 180 degrees. In contrast, BrC and OC's $F_{11}$ initially decrease then increase with a local minimum at a scattering angle of 130 degrees. The curves of $-F_{12}/F_{11}$ for BC aerosols are single-peaked. A positive-valued peak appears at a scattering angle near 90 degrees. For BrC or OC aerosols, a negative-valued peak has also been observed at about 150 degrees in addition to the positive peaks. These two characteristics hold for the entire particle-size spectrum in this study. The non-sphericity of BC particles, which results in different angular dependency curve for spherical particles in $F_{33}$ and $F_{34}$, have not been covered in this paper.

We applied the optical parameters of different types of aerosols mentioned above into the SOS algorithm to calculate BRF, pBRF, and DoLP at the TOA. The atmosphere-ocean interface was assumed to be rough because of the wind flowing over it. The angular dependency of BRF and DoLP follow the similar changing patterns of $F_{11}$ and $-F_{12}/F_{11}$, respectively. The angular dependencies of pBRF are mainly determined by Rayleigh scattering of the atmosphere because of the low magnitudes of the polarized phase functions ($F_{12}$) of particles. The comparison of the reflectance factors between strongly and weakly light-absorbing aerosols indicates their different levels of sensitivities to effective diameter $d_{\text{eff}}$. Fluctuations in the values of the particle imaginary refractive index $RI_i$ and $d_{\text{eff}}$ within in the low $RI_i$ region will more effectively influence the magnitudes of BRF and pBRF. DoLP, which is defined as the ratio of pBRF and BRF, amplifies the sensitivity to both $RI_i$ and $d_{\text{eff}}$. In this study, DoLP ranges widely from 0.23 to 0.88 at a scattering angle close to 90 degrees. We found this parameter to be most significantly affected by the $RI_i$ values of the particles, especially when the values are low as found in BrC aerosols. Quantitative sensitivity analysis shows particle sizes influence the level of DoLP's sensitivity to $RI_i$. Our results indicate that polarization data contain more particle-size information when $RI_i$ is less than 0.4. The spectral dependency of DoLP within the bandwidth of 470 nm proves that the reliability of polarimetric measurement is highly sensitive to both particle $RI_i$ values and size distributions. The next step of our research is to quantify the aerosol microphysical properties' sensitivity over anisotropic surfaces with the dataset of the bidirectional reflectance distribution function (BRDF) and bidirectional polarization distribution function (BPDF). We will also address the question of retrievability of these key aerosol properties with multi-spectral as well as multi-angle photo-polarimetric data.

**ACKNOWLEDGEMENT**

Portions of this work were performed at the Jet Propulsion Laboratory, California Institute of Technology, under a contract with the National Aeronautics and Space Administration (NASA). RC and CZ acknowledge support from the US National Science Foundation (AGS-1455215 and AGS-1926817) and the NASA ACCDAM program (NNH20ZDA001N). CZ would like to acknowledge partial support received from the McDonnell International Scholars Academy at Washington University in St. Louis.

**Table 1**. Effective diameter, $d_{eff}$, and their associated geometric mean radii, $r_g$, for the range of particle sizes in this study.

|  | Size 1 | Size 2 | Size 3 | Size 4 | Size 5 | Size 6 | Size 7 |
|---|---|---|---|---|---|---|---|
| $r_g$ (nm) | 11.3 | 22.5 | 44.9 | 90.1 | 179.9 | 359.8 | 719.6 |
| $d_{eff}$ (nm) | 39.1 | 78.2 | 156.0 | 313.0 | 625.0 | 1250.0 | 2500.0 |

**Table 2.** Refractive indices for simulated particle models. Black carbon ($BC_{high}$ and $BC_{low}$) and organic carbon (OC) have spectrally independent refractive indices across the 350nm – 885nm wavelength range. The refractive index of brown carbon (BrC) has a real part of 1.60, and the imaginary part of the refractive index follows the Kramers–Kronig relation between 350 nm – 572 nm, with constants of $a = 10^{29}$ s$^{-2}$, $\gamma = 2 \times 10^{13}$ s$^{-1}$, $v = c/\lambda$, $\lambda_0 = 300$ nm. At $\lambda$ between 572 nm – 885 nm, BrC has $n = 1.60 + 0.001i$.

| $\lambda$ [nm] | $BC_{high}$ | | $BC_{low}$ | | BrC | | OC | |
|---|---|---|---|---|---|---|---|---|
|  | $RI_r$ | $RI_i$ | $RI_r$ | $RI_i$ | $RI_r$ | $RI_i$ | $RI_r$ | $RI_i$ |
| **350~572** | 1.95 | 0.79 | 1.75 | 0.453 | 1.60 | $a\dfrac{\gamma v}{(v_0^2 - v^2)^2 + (\gamma v)^2}$ | 1.53 | 0.001 |
| **572~935** | 1.95 | 0.79 | 1.75 | 0.453 | 1.60 | 0.001 | 1.53 | 0.001 |

**Table 3.** Viewing geometries for multiple forward simulations

| View number | View zenith angle (deg) | Relative azimuth (deg) | Scattering angle (deg) | Solar zenith angle (deg) |
|---|---|---|---|---|
| 1 | 68.38 | 0 | 81.62 | 30.0 |
| 2 | 61.19 | 0 | 88.81 | 30.0 |
| 3 | 50.19 | 0 | 99.81 | 30.0 |
| 4 | 50.19 | 60 | 111.25 | 30.0 |
| 5 | 32.07 | 60 | 126.95 | 30.0 |
| 6 | 9.00 | 60 | 144.71 | 30.0 |
| 7 | 9.00 | 120 | 153.44 | 30.0 |
| 8 | 32.07 | 180 | 177.93 | 30.0 |

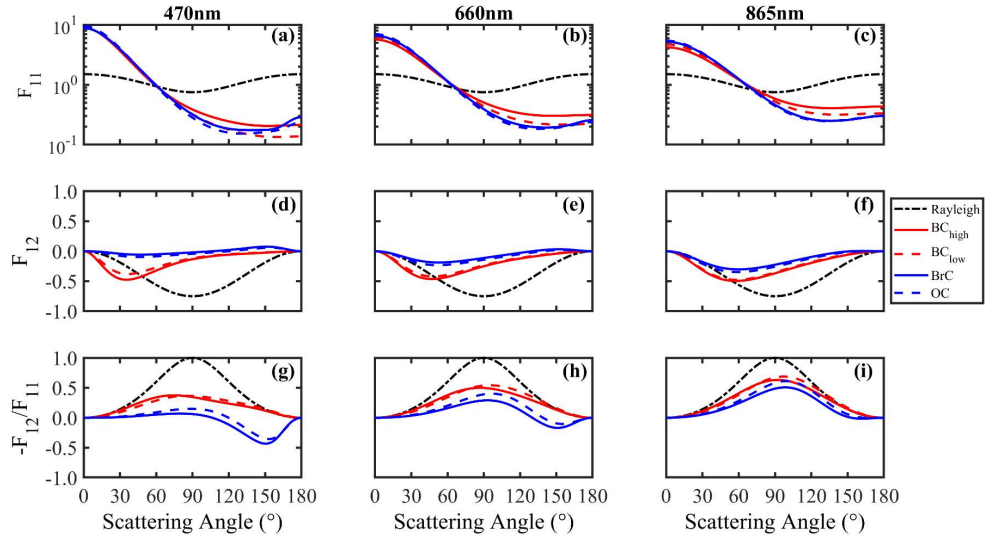

**Figure 1.** Scattering phase matrix elements $F_{11}$ (Fig. 1 (a), (b), and (c)), $F_{12}$, and $-F_{12}/F_{11}$ of the four aerosol types at $\lambda = 470$, 660, and 865nm. The particle sizes follow a log-normal distribution with an effective particle diameter $d_{eff} = 90$ nm and geometric width $\sigma_g = 1.6$. The Rayleigh scattering by atmosphere is shown as the dashed-dotted black line, $BC_{high}$ as the solid red lines, $BC_{low}$ as the dashed red lines, BrC as the solid blue lines, and OC as the dashed blue lines.

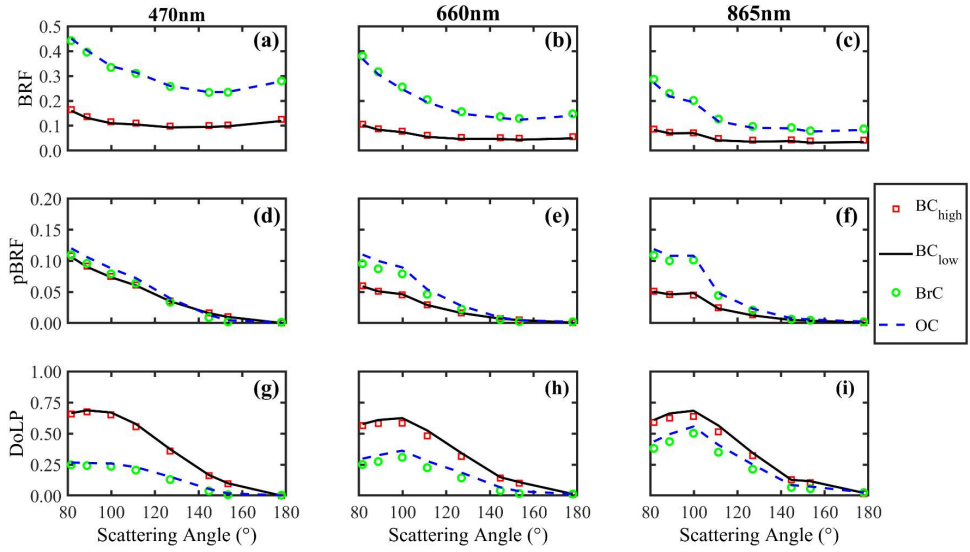

**Figure 2.** BRF, pBRF, and DoLP of the four aerosol types at three studied wavelengths. The particle sizes follow a log-normal distribution with an effective diameter of $d_{\text{eff}}$ = 90 nm and a geometric width of $\sigma_g$ = 1.6.

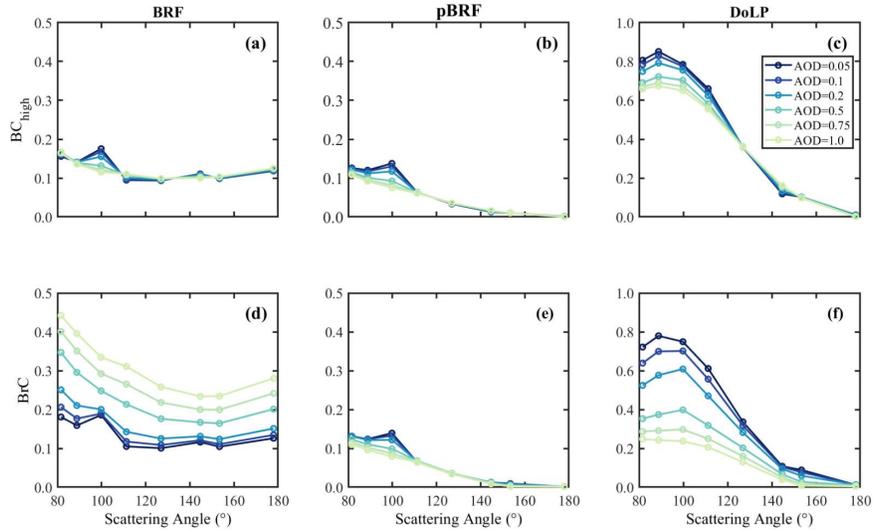

**Figure 3.** The influences of boundary layer AOD on the reflectance factors of BChigh and BrC particles. The particle sizes follow a log-normal distribution with an effective diameter of $d_{\text{eff}}$ = 90 nm and a geometric width of $\sigma_g$ = 1.6.

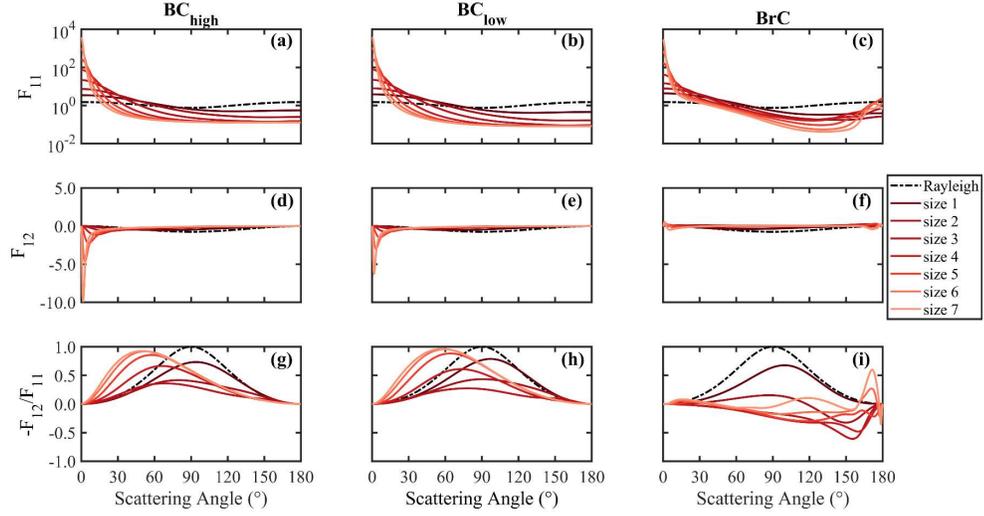

**Figure 4.** The scattering phase matrix elements $F_{11}$, $F_{12}$, and $-F_{12}/F_{11}$ of three types of aerosols with different effective diameters at $\lambda = 470$nm. The particle sizes follow log-normal distributions with a fixed geometric width of $\sigma_g = 1.6$ and 7 different effective diameters detailed in Table 1. The Rayleigh background is shown by the dashed-dotted black line.

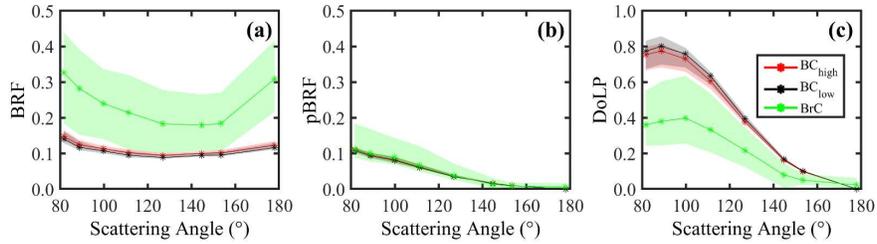

**Figure 5.** BRF, pBRF, and DoLP of $BC_{high}$, $BC_{low}$, and BrC aerosols at $\lambda = 470$ nm. The solid lines indicate the mean values of the reflectance factors. The width of the shaded area demonstrates the value ranges of the three reflectance factors as the mean effective diameters $d_{eff}$ increase from 39.1 nm to 2500.0 nm.

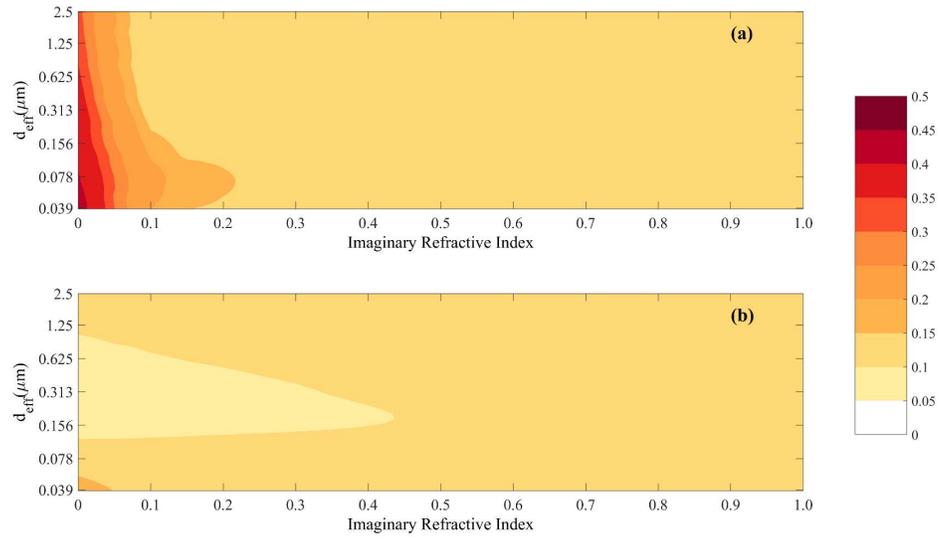

**Figure 6.** Contour plots of the (a) BRF and (b) pBRF at a scattering angle of 81.6 degrees and incident wavelength of 470nm. The magnitudes of the reflectance factors are shown in the sequential color scheme.

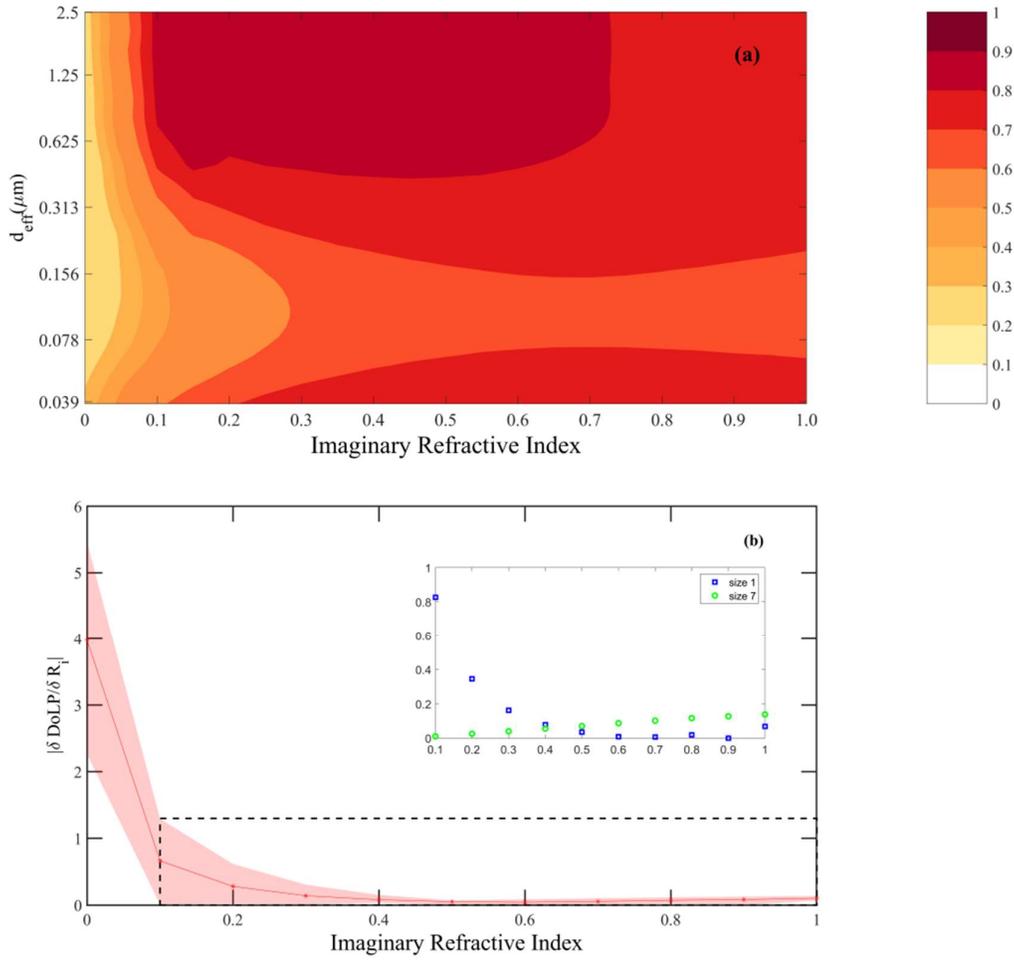

**Figure 7.** (a) DoLP at a scattering angle of 81.6 degrees and incident wavelength of 470nm. The magnitudes of DoLP are shown in the sequential color scheme. (b) Moving average of Jacobians for DoLP with respect to the imaginary refractive index, $R_i$. The red solid line shows the mean Jacobian of DoLP for the aerosols which follow the seven studied size distributions. The inset figure shows Jacobians of DoLP for size 1 and 7 aerosols when $R_i$ exceeds 0.1 (The area bounded by dashed-line box).

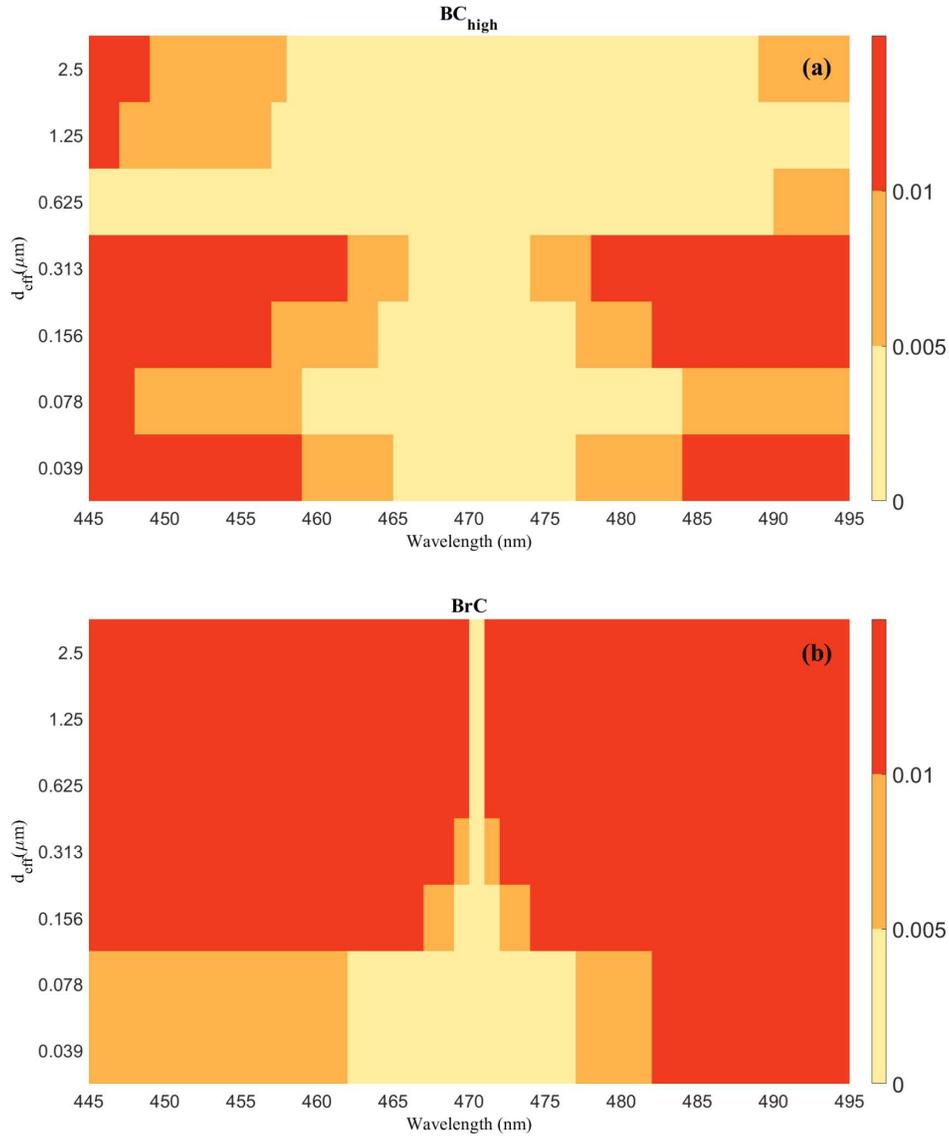

**Figure 8.** Relative deviation of DoLP for (a) $BC_{high}$ and (b) BrC particles within the 45-nm bandwidth of 470 nm. The particle sizes follow a log-normal distribution with a geometric width of $\sigma_g = 1.6$.

Calibration and validation of airborne multiangle spectropolarimetric imager (AirMSPI) polarization measurements. Appl Opt. 2018;57:4499-513.

# Polarimetric Sensitivity of Light-Absorbing Carbonaceous Aerosols Over Ocean: A Theoretical Assessment

Chenchong Zhang[a], William R. Heinson[a,b], Michael J. Garay[c], Olga Kalashnikova[c], Rajan K. Chakrabarty[a,*]

[a] *Center for Aerosol Science and Engineering, Department of Energy, Environmental and Chemical Engineering, Washington University in St. Louis, St. Louis, MO 63130, USA*
[b]*Earth System Science Interdisciplinary Center (ESSIC), University of Maryland, College Park, MD 20740 USA, and Climate and Radiation Laboratory, NASA Goddard Space Flight Center, Greenbelt, MD, USA*
[c]*Jet Propulsion Laboratory, California Institute of Technology, Pasadena, CA 91109, USA*

*Corresponding author.
E-mail address: chakrabarty@wustl.edu

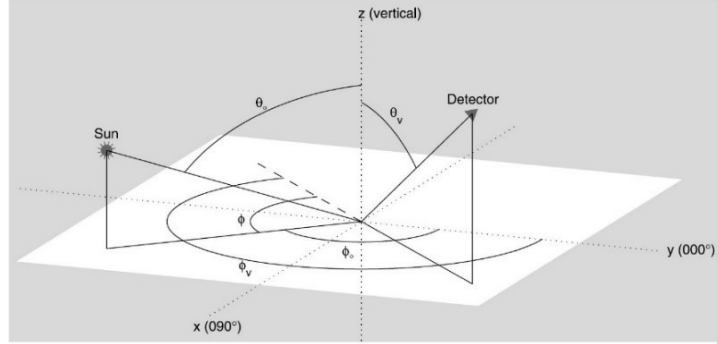

Figure S1. The schematic of the solar zenith angle $\theta_0$, viewing zenith angle $\theta_v$, solar azimuth angle $\phi_0$, viewing azimuth angle $\phi_v$, and relative azimuth angle $\phi$ (adapted from Hudson et al., 2006 [32]).

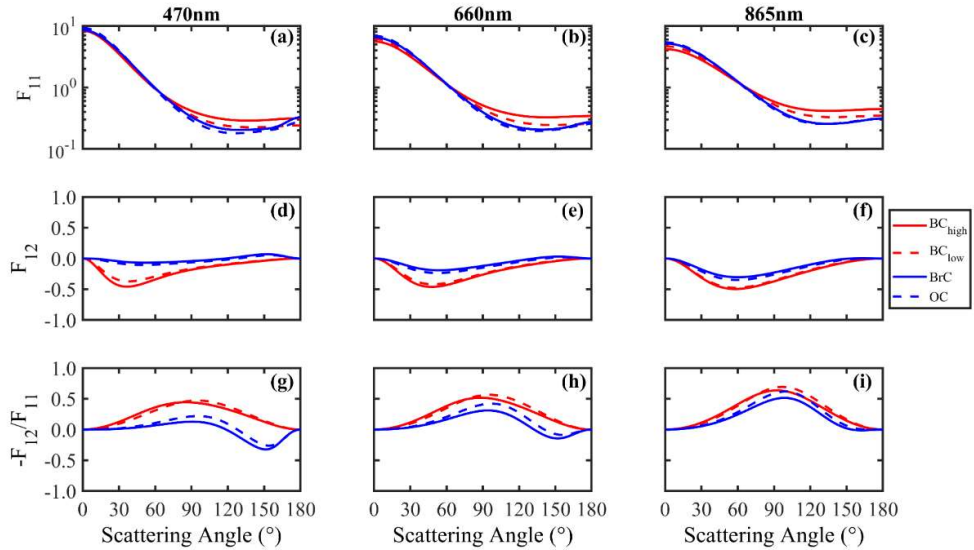

Figure S2. Scattering phase matrix elements $F_{11}$ (Fig. S2 (a), (b), and (c)), $F_{12}$, and -$F_{12}/F_{11}$ of the four aerosol types mixed with Rayleigh atmosphere at $\lambda$ = 470, 660, and 865nm. The particle sizes follow a log-normal distribution with an effective particle diameter $d_{\text{eff}}$ = 90 nm and geometric width $\sigma_g$ = 1.6.

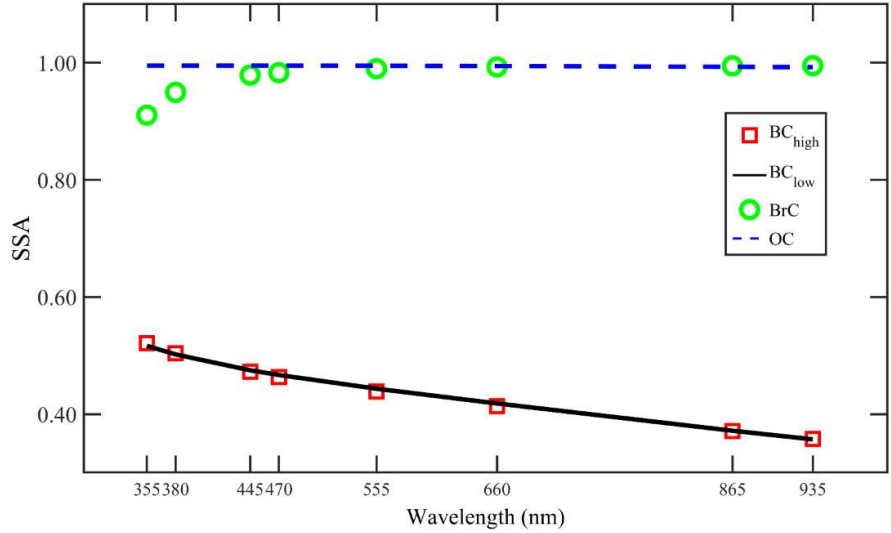

Figure S3. The spectral dependency of SSA for four types of aerosols at all eight wavelength bands of AirMSPI. The particle sizes follow a log-normal distribution with an effective diameter of $d_{eff}$ = 90 nm and a geometric width of $\sigma_g$ = 1.6

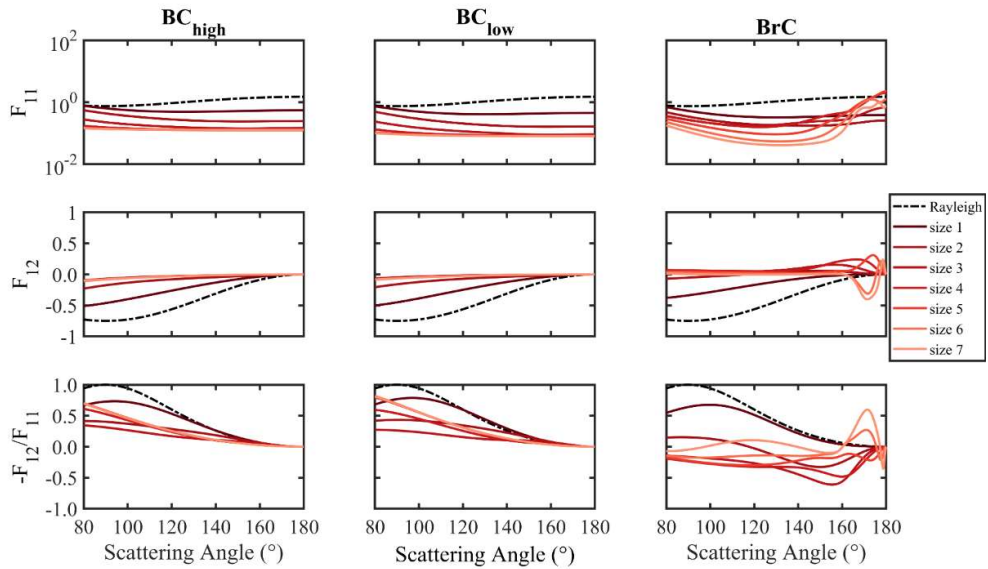

Figure S4. The scattering phase matrix elements $F_{11}$, $F_{12}$, and $-F_{12}/F_{11}$ of $BC_{high}$, $BC_{low}$, and BrC with different effective diameters at $\lambda$ = 470nm. The particle sizes follow the same manner as Fig. 3. In this figure, we highlight the angular dependencies of the phase matrix values in a scattering range between 80 to 180 degrees.

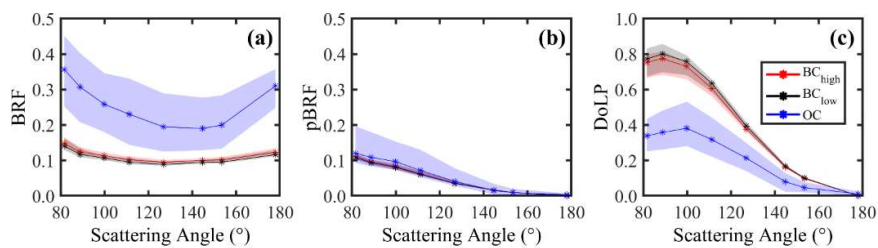

Figure S5. BRF, pBRF, and DoLP of $BC_{high}$, $BC_{low}$, and OC aerosols at $\lambda = 470$nm. The solid lines indicate the mean values of the reflectance factors. The width of the shaded area demonstrates the value ranges of the three reflectance factors as the mean effective diameters $d_{eff}$ increase from 39.1 nm to 2500.0 nm.

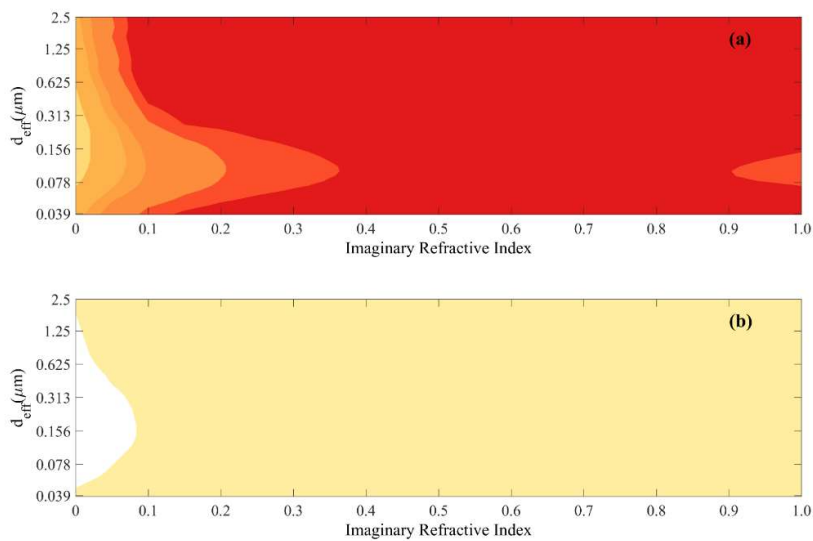

Figure S6. Contour plots of the DoLP values at a scattering angle of (a) 99.81 and (b) 144.71 degrees. The wavelength of incident light is fixed at 470nm. The magnitudes of the reflectance factors are shown in the sequential color scheme.